# When Incentives and Nudges Meet:
# Promoting Budget Allocations for Undervalued Policies

Makoto Kuroki[1]    Shusaku Sasaki[2]

Date of Written: 13th May 2025

## Abstract

Budget officers often assess public project proposals based on available financial support and expected outcomes. However, behavioral factors such as time discounting and psychological hesitation may lead to underinvestment in programs with delayed but significant benefits. This study investigates whether financial incentives and non-financial nudges can influence budgetary decisions in local governments. We conducted a nationwide mail-based survey experiment targeting budget officers in Japanese municipalities and received 490 valid responses. Using a 2×2 randomized design, we tested the independent and combined effects of a financial incentive (a 50% national subsidy) and a non-financial nudge (loss framing and peer information). All three treatments significantly increased assessed budget amounts compared to the control group. The largest effect appeared in the combination group (approximately 1.1 million JPY higher, $p < .01$). Both the financial incentive and the nudge independently increased assessments by approximately 650,000–670,000 JPY ($p < .01$). Notably, only the nudge raised the self-financed portion of the budget. These findings demonstrate how low-cost behavioral interventions can improve budget assessments for undervalued projects. As an application, we embedded the experiment in a case involving HPV vaccine promotion in Japan, but the approach can be broadly relevant to other public policy domains requiring forward-looking budget decisions.

[1] Yokohama City University; kuroki@yokohama-cu.ac.jp
[2] The University of Osaka; ssasaki.econ@cider.osaka-u.ac.jp

The authors are listed in alphabetical order. The first two authors contributed equally to this work and share first authorship. This study is financially supported by the Japan Society for the Promotion of Science [grant number: 21H00762, 23KK0232, 24K00264, 25H00388]. We obtained approval for this study from the ethics review committees of Yokohama City University Kanazawa Hakkei Campus (Hachi 2023-14) and the Center for Infectious Disease Education and Research at Osaka University (2023CRER1010). We also pre-registered the overview, design, and procedure plan for the survey experiment with the AEA RCT Registry (Kuroki & Sasaki, 2023b).

# 1. Introduction

Understanding the decision-making processes underlying budget assessments has been a central topic in the fields of public administration and public sector accounting (Anessi-Pessina et al., 2016). However, these processes have long been considered a black box, as most previous studies have relied on descriptive surveys asking what kinds of information are used in assessments or on correlation analyses examining relationships between assessed amounts and the existence of processes or output indicators (Liguori et al., 2012; Marti, 2013; Melkers and Willoughby, 2001, 2005; Zaltsman, 2009). Nevertheless, building on these surveys, analyses, and related theories, it has been suggested that budget officers in both central and local governments make budgetary decisions regarding public projects based on a range of available information, including projected policy outcomes (Baekgaard, 2015; Demaj, 2017; Nielsen & Baekgaard, 2015), as well as financial and non-financial performance data (Gilmour & Lewis, 2006; Heinrich, 2012; Kuroki, 2022; Reck, 2001).

Although expected outcomes constitute one of the most important sources of information in budget assessments, they are frequently underestimated (Jacobs, 2011, 2016). This tendency has been attributed to a well-documented behavioral bias, whereby individuals systematically discount future benefits relative to present ones (Loewenstein & Prelec, 1992). As a consequence, policy programs that entail short-term costs but deliver long-term societal benefits—such as those in environmental, health, and education sectors—are often assessed less favorably in the budgeting process (Jacobs, 2011, 2016; McGowan et al., 2022). Addressing the tendency to undervalue such projects has been widely recognized as a pressing issue in both academic and policy domains.

Two main approaches have been discussed in the literature to address this issue. The first, which has traditionally received greater emphasis, focuses on financial information and cost-related strategies. For example, reducing the administrative costs of a project may improve its cost-efficiency, thereby increasing the likelihood of its selection even when future outcomes are discounted (Anessi-Pessina et al., 2016; Mauro, 2021; Kuroki & Motokawa, 2022). A typical scheme associated with this approach is the use of subsidy incentives from central governments to local governments. This scheme, commonly referred to as a matching fund, involves central government subsidies covering a portion of project costs, thus reducing the fiscal burden on local governments and potentially enhancing the attractiveness of the project.

Second, in recent years, growing attention has been paid to strategies that emphasize non-financial information. This trend corresponds with Simon's (1997) concept of bounded rationality in budgetary decision-making. Rubin (2008) also underscored the significance of individual decision-making in budget assessments, encouraging further research into how various individual decision-making characteristics affect budgetary outcomes (Mohr & Kearney, 2021). One promising strategy that incorporates both bounded rationality and individual-level decision-making is the use of nudges in behavioral economics. In a recent field-based survey experiment, Kuroki and Sasaki (2023a) found that presenting expected project outcomes in a loss-framing format or providing social comparison information based on other local governments' budget assessments led to higher evaluated budget amounts.

Previous academic research has primarily examined the effects of financial and non-financial information separately, leaving the comparative effectiveness of these approaches unclear. While monetary interventions—such as matching funds—are generally assumed to have a strong influence on budget decisions, they impose substantial fiscal burdens on central governments in the form of subsidies. In contrast, non-monetary interventions, such as nudges, are relatively low-cost (Thaler & Sunstein, 2009; Sunstein, 2014); however, their effectiveness is highly dependent on the characteristics and surrounding conditions of the target population, and their average impact may be modest (DellaVigna & Linos, 2022). These observations raise important questions about which type of intervention yields greater cost-effectiveness. Moreover, in actual budget assessments, financial and non-financial information are frequently considered in tandem, yet the interaction between the two remains largely unexplored (Mauro, 2017). Accordingly, the potential effects of integrated strategies that target both types of information have not been sufficiently investigated.

Based on the above discussion, we formulate the following three research questions:

*RQ1: What is the effect of financial incentives and non-financial nudges on budget assessments?*
*RQ2: How do the effects of financial incentives and non-financial nudges differ?*
*RQ3: What is the effect of combining financial incentives with non-financial nudges on budget assessments?*

To address these questions, we conducted a mail-based survey experiment between November and December 2023, targeting budget officers in 1,741 municipalities across Japan. In the experiment, respondents were presented with a hypothetical scenario describing a public awareness campaign concerning the resumption of governmental recommendations for HPV (Human Papillomavirus) vaccination in Japan, which had been suspended from June 2013 to October 2021. Respondents were then asked to assess the project and indicate the amount of budget they would allocate.

To investigate this issue, we randomly assigned municipalities into four groups and implemented a 2×2 between-subjects experimental design to examine the effects of financial incentives and non-financial nudges. We classified the groups as follows: (1) no intervention, (2) financial incentive only, (3) non-financial nudge only, and (4) both financial incentive and non-financial nudge. The financial incentive consisted of a scheme in which the national government would subsidize 50% of the project cost. The non-financial nudge included a message informing participants that other municipalities had already resumed the vaccination campaign, combined with a loss-framing description of the expected future outcomes of doing so.

At the time of the survey and the writing of this paper, the central government had not introduced any subsidy schemes to support local governments in implementing HPV (Human Papillomavirus) vaccination awareness campaigns. However, since the target vaccination coverage set by the Ministry of Health, Labour and Welfare had not yet been achieved, we considered it highly relevant from a policy perspective to examine the effects of financial incentives, such as subsidies. In addition, as discussed in Section 2, the historical context surrounding HPV vaccination in Japan suggests the existence of psychological barriers that may influence budgetary decision-making for such campaigns. Therefore, we deemed it appropriate to address our research questions in the context of the HPV vaccination recommendation program in Japan.

Our mail-based survey yielded valid responses from budget officers in 490 municipalities, corresponding to a valid response rate of 28.14%. The results of the statistical analysis show that the total assessed budget amount—including the subsidy—was highest among the group that received both the financial incentive and the non-financial nudge, with an estimated increase of approximately 1,110,000 Japanese yen (JPY) compared to the control group ($p < .01$). When the financial incentive or the non-financial nudge was provided independently, the assessed budget amount was also significantly higher than that of the control group ($p < .01$), with estimated increases of approximately 650,000 to 670,000 JPY, respectively. These findings suggest that non-financial nudges can raise assessed budget amounts—which tend to be underestimated—to a level comparable to that achieved by financial incentives such as subsidies. Furthermore, combining both interventions may generate an additive or amplifying effect.

## 2. Background on HPV Vaccination

Japan operates under a two-tier administrative system consisting of the central government and local governments (prefectures and municipalities), with local allocation tax grants transferred from the central government to support local finances. In the context of vaccination policy, the central government—particularly the Ministry of Health, Labour and Welfare—determines which vaccines are covered by public insurance, while municipalities are responsible for developing implementation frameworks and conducting public awareness campaigns. For example, municipalities conduct surveys of hospitals and clinics that offer vaccinations and provide subsidies to create environments conducive to vaccine uptake.

With respect to HPV (Human Papillomavirus) vaccination, local governments began subsidizing the cost of administering the bivalent vaccine to girls aged 13 to 16 in 2010, following a recommendation by the central government. By 2013, both the bivalent and quadrivalent vaccines had been incorporated into Japan's routine immunization program for girls aged 12 to 16. However, in June 2013, the Ministry of Health, Labour and Welfare suspended its active recommendation for the HPV vaccine, citing the need to further assess vaccine safety. This decision was influenced by widespread media reports of adverse events following vaccination, which heightened public concern. Consequently, vaccination rates fell sharply—from approximately 70% of eligible girls to below 1%.

In April 2017, a research team under the Ministry of Health, Labour and Welfare published an epidemiological study showing that adverse events reported following HPV vaccination were also observed among unvaccinated girls in the same age group. Based on this finding, the government concluded that the adverse events were not causally linked to the vaccine. As a result, the government resumed active recommendations in November 2021. From April 2022, eligible girls were able to receive the vaccine free of charge as part of the routine immunization schedule. In addition, women born between 1997 and 2005 who missed the opportunity for routine vaccination became eligible for free catch-up vaccination between April 2022 and March 2025.

Nevertheless, despite the resumption of official recommendations, vaccination uptake has not increased as expected. Projections indicate that, if current trends continue, the cumulative vaccination rate will plateau at 43.16% (Yagi et al., 2024). This rate is substantially below the WHO's 90% target and the 70% coverage prior to the suspension, highlighting the need for additional efforts to improve uptake. While vaccine costs for eligible girls are covered by national funds, the cost of awareness campaigns is often borne by local governments, making the implementation of additional measures highly dependent on local government budgets.

The budgeting process for vaccination programs in Japanese municipalities is similar to those observed in other countries such as Germany, the United Kingdom, and France. In this process, health departments that oversee vaccine administration submit budget requests to the finance departments, which assess these requests in accordance with policy guidance from the central government. Individual budget officers conduct the initial evaluation of the proposed budgets, followed by review by senior officials such as department heads. These reviews result in a secondary budget proposal, which is then approved by the head of the local government and submitted to the municipal assembly for final approval. Although the case of HPV vaccination recommendations is specific to Japan, the structure and process of local government budgeting exhibit notable similarities with those of many other countries.

## 3. Hypothesis Setting
### 3.1. Financial Incentive Only
In this experiment, we use a financial incentive in which half of the cost associated with implementing the awareness campaign would be covered by a central government subsidy. Many local governments in Japan operate under severe fiscal constraints due to the increasing burden of elderly care expenditures, which makes them reluctant to undertake new policy initiatives using only their own financial resources. As a result, the central government occasionally provides fiscal support, such as the incentive described above, to promote smoother policy implementation at the local level.

In the context of budget assessments for the HPV vaccination awareness campaign, if the primary constraint faced by local finance departments is the cost, then a subsidy that reduces the financial burden on local governments is expected to increase the likelihood that they will allocate a non-zero budget. However, the total assessed budget—which includes both the local government's own contribution and the subsidy—ultimately depends on the local government's willingness to allocate its own resources. If the finance department decides to maintain or even increase the self-financed budget despite the subsidy covering half the cost, the incentive would lead to a net increase in the total budget amount. Conversely, if the finance department offsets its own contribution by an amount equivalent to the subsidy, the total assessed budget would remain unchanged.

A similar mechanism has been discussed in the literature on charitable giving, where matching schemes can theoretically yield two distinct outcomes: an increase in the total donation amount or no change at all. Empirically, the former outcome has been observed more frequently (Epperson & Reif, 2019; Karlan & List, 2007). In this setting, whether the financial incentive leads to an increase in the total assessed budget remains an empirical question.

### 3.2. Non-Financial Nudge Only

In this experiment, we introduce a non-financial nudge by informing participants that other municipalities had already resumed vaccination recommendations and by emphasizing the anticipated future outcomes of doing so through a loss-framed message. If the primary constraint for local finance departments in assessing the budget for the awareness campaign lies in psychological hesitation, this nudge is expected to be effective. For instance, due to the prolonged suspension of active HPV vaccination recommendations in Japan, finance departments may remain uncertain about whether allocating resources to the campaign will produce the intended effects. Finance departments further may experience psychological resistance to acting as the first mover, when neighboring municipalities have not yet resumed their campaigns. If such psychological bottlenecks are indeed significant, the nudge described above could contribute to an increase in the assessed budget amount.

### 3.3. Combination

We argue that the effect of combining financial incentives with non-financial nudges depends on the proportion of finance departments that determine their assessed budget amounts according to the mechanism outlined in Hypothesis 1 versus Hypothesis 2. If nearly all departments follow the logic of Hypothesis 1, the combined intervention will yield the same outcome as the financial incentive alone, with the nudge having no additional effect. Conversely, if nearly all departments operate under Hypothesis 2, the combined intervention will replicate the effect of the nudge alone, rendering the financial incentive ineffective.

In scenarios where both types of departments exist in significant proportions, the combined intervention will impact both groups. However, the overall effect on the assessed amount becomes more complex and contingent on whether the incentive alone increases the assessed amount. If the incentive does not influence the total assessed amount, the combined intervention will only increase the assessed amount for departments operating under mechanism 2, making the combined effect equivalent to the effect of the nudge alone, with the incentive having no effect. Conversely, if the incentive does result in an increased assessed amount, the combined intervention will elevate the assessed amount across both types of departments, resulting in a combined effect that exceeds the impact of either the incentive or the nudge alone.

## 4. Survey Experiment Design

### 4.1. Field-Based Survey Experiment

In previous research, researchers have relied on descriptive surveys asking what kinds of information are used in assessments or on correlation analyses examining relationships between assessed amounts and the existence of processes or output indicators (Liguori et al., 2012; Marti, 2013; Melkers and Willoughby, 2001, 2005; Zaltsman, 2009). However, these approaches have often proven insufficient in revealing the actual decision-making processes of budget officers.

To address this limitation, we employ a field-based survey experiment. This method involves constructing multiple hypothetical scenarios that vary in terms of the information and conditions provided, and randomly assigning them to participants. By comparing the budgetary decisions made across different experimental groups, we identify the effects of specific types of information and contextual factors on decision-making.

Field-based survey experiments have been widely adopted in fields such as political science and public administration. By involving real-world decision-makers as subjects and designing scenarios that reflect plausible policy contexts, this method could enhance the generalizability of experimental findings. It also addresses the limitations of laboratory experiments—which often rely on student samples and suffer from limited generalizability—as well as field experiments, which are frequently constrained by a lack of data availability (Druckman, 2014).

### 4.2. Content of Survey Experiment

We describe the hypothetical scenario used in the questionnaire in detail in the **Appendix 1**. In this scenario, respondents—who were real budget officers—were instructed to assume the role of officials in a municipality with a population of 50,000. They were presented with a case in which the health department requested a 5 million JPY budget for a new public awareness campaign related to the resumption of HPV vaccination recommendations. Based on the reference information provided, respondents were asked to determine the amount they would allocate to the campaign.

To reflect the general tendency for strict assessments for new projects, we explicitly stated in the scenario that the municipality was facing severe financial constraints and that the finance department had recommended capping the requested amount at the level of the previous year's assessed budget. All groups were provided with common baseline information, including the previous year's budget, which was set at 0 JPY, as the request pertained to the period before the resumption of active recommendations. We also included performance information such as the HPV vaccination rates in the recent years, projected coverage for the current fiscal year, and the target rate for the project year.

To test the hypotheses presented in Section 3, we implemented a 2×2 between-subject design that varied the provision of financial incentives and non-financial nudges. Respondents were randomly assigned to one of four groups: no intervention, financial incentive only, non-financial nudge only, and the combination. The financial incentive consisted of a subsidy scheme whereby the central government would cover half the cost of the awareness campaign. The non-financial nudge consisted of a message informing participants that other municipalities had already resumed vaccination recommendations and highlighting the anticipated future outcomes using a loss-framed message. For the three intervention groups, the following messages were added to the common information described above.

**Financial Incentive Group (denoted as "*Incentive*" in the tables and figure):**
"This year, half of the cost for this awareness campaign will be funded by central government subsidies."

**Non-Financial Nudge Group (*denoted as "Nudge" in the tables and figure*):**
"Neighboring municipalities have already launched similar awareness campaigns, and the vaccination rate for 2023 is projected to reach 20%, drawing increased public attention. If this awareness campaign is not implemented, the vaccination rate in this municipality is expected to remain low."

**Combination Group (*denoted as "Incentive×Nudge" in the tables and figure*):**
"This year, half of the cost for this awareness campaign will be funded by central government subsidies. Neighboring municipalities have already launched similar awareness campaigns, and the vaccination rate for 2023 is projected to reach 20%, drawing increased public attention. If this awareness campaign is not implemented, the vaccination rate in this municipality is expected to remain low."

Before implementing the survey experiment, we conducted multiple interviews with researchers specializing in public health and pediatric medicine to identify the most appropriate theme for our study within the scope of healthcare policy. While we considered several potential areas—including promoting awareness of childhood vaccination, strengthening emergency and healthcare service systems, and improving healthcare infrastructure—we concluded that leveraging the timing of the survey, which coincided with low HPV vaccination rates, made HPV-related awareness the most suitable topic.

We also conducted interviews with six administrative managers from four local governments to assess the validity of the hypothetical scenarios. We discussed whether it would be preferable to frame the scenario around a generic vaccine or to specify a particular one, such as the HPV vaccine. In addition, we examined whether a matching fund structure—in which the central government covers half of the project cost—together with the nudge messages would be perceived as realistic. The feedback suggested that specifying an actual vaccine made the scenario easier to evaluate, that a 50:50 cost-sharing scheme between central and local governments was common practice, and that the explanation of the nudge message was clear and contextually appropriate. Based on this feedback, we made minor adjustments to enhance the realism of the scenario and finalized the experimental materials.

In the response section for the assessed amount, we provided space for each budget officer to record their individual assessment, along with a separate field to indicate the amount determined after consulting with their superiors, such as managers. To measure the degree to which various types of information were referenced during the assessment process, we included a 7-point Likert scale question. We also collected information on the attributes of the responding budget officers and added an open-ended section for them to explain the reasoning behind their assessed amount.

We obtained approval for this survey experiment from the ethics review committees of Yokohama City University Kanazawa Hakkei Campus (Hachi 2023-14) and the Center for Infectious Disease Education and Research at Osaka University (2023CRER1010). We also pre-registered the overview, design, and procedure plan for the survey experiment with the AEA RCT Registry (Kuroki & Sasaki, 2023b).

### 4.3. Randomization

Japan comprises 1,741 municipalities, including designated cities, general cities, towns, and villages. Designated cities typically have populations of over 500,000, general cities generally exceed 10,000 residents, and towns and villages usually have fewer than 10,000 residents. However, exceptions exist due to historical classifications, with some towns exceeding 10,000 and some cities having fewer than 10,000 residents. Tokyo, as a special metropolitan area, is subdivided into 23 special wards (*ku*). We conducted randomization within each municipal stratum and distributed four types of questionnaires accordingly.

[Table 1 is here.]

On November 13, 2023, we sent survey questionnaires to budget officers in the finance departments of all 1,741 municipalities, requesting responses by December 10, 2023. As shown in **Table 1**, Panel B, we received responses from 558 municipalities. After excluding responses with missing data, we retained 490 for analysis. A chi-square test confirmed no statistically significant differences in response rates among the randomly assigned groups by strata, indicating that the data collection process was free from systematic bias.[3]

In our regression analysis, we included control variables based on previous studies that have identified factors affecting budget assessments and vaccination coverage. The control variables included the respondent's age, gender, years of experience in the department, number of departmental staff, municipality population size, the proportion of girls aged 12–16, population density, net population change, fiscal strength index, and average taxable income. We collected information on individual and departmental attributes from the questionnaire and obtained municipal-level data from e-Stat, the Government of Japan's public statistics open data platform.

---

[3] Although no statistically significant differences were observed, the number of responses from the control group was slightly higher than those from the other groups. To address this, we conducted an additional analysis in which we randomly reduced the sample size of the control group to match that of the other groups. However, since the results remained consistent, we present the analysis using all 490 samples in this paper.

## 5. Results
### 5.1. Descriptive Results

[Table 2 is here.]

**Table 2**, Panel A presents descriptive statistics for the 490 budget officers in our sample, including the assessed amounts, the proportion of non-zero assessed amounts, individual attributes of the assessors, and demographic characteristics of their respective municipalities. The mean (median) assessed amount was 1,346,330 JPY (200,000 JPY), suggesting that the assessments were relatively conservative compared to the requested amount of 5,000,000 JPY. A total of 63.5% of respondents entered an assessed amount of at least 1 JPY. The majority of the respondents were male (86.9%) and had worked in their current department for an average of 3.3 years.

**Table 2**, Panel B reports descriptive statistics by treatment group. First, we confirmed that the covariates did not differ significantly across groups, indicating successful randomization and balanced group composition. The mean assessed amount was highest in the combination group (1,872,460 JPY), and the differences between this group and the financial incentive group (1,436,070 JPY, $p < 0.10$), the non-financial nudge group (1,410,300 JPY, $p < 0.10$), and the control group (757,244 JPY, $p < 0.01$) were all statistically significant. We also observed statistically significant differences between the financial incentive group and the control group ($p < 0.01$), as well as between the non-financial nudge group and the control group ($p < 0.01$). However, there was no statistically significant difference between the financial incentive group and the non-financial nudge group. Accordingly, the assessed amounts followed the pattern: combination group > financial incentive group ≈ non-financial nudge group > control group.

The proportion of respondents who entered an assessed amount of at least 1 JPY also followed the same ordering. Specifically, the proportion was 80.5% in the combination group, 64.9% in the financial incentive group, 62.3% in the non-financial nudge group, and 48.5% in the control group.[4]

---

[4] In the setting of our scenario, the assessed amount in the previous fiscal year was zero. However, in the current fiscal year, even within the control group, 48.5% of respondents selected a non-zero assessed amount, and the average assessed amount reached 757,244 JPY. This shift appears reasonable if we consider that the previous year marked the immediate aftermath of the resumption of active HPV vaccination recommendations, during which many municipalities were still hesitant. By contrast, in the current fiscal year—more than one year after the resumption—some municipalities may have started to consider allocating budgets for the campaign, making such changes plausible.

**[Figure 1 is here.]**

**Figure 1** displays the assessed budget amounts by treatment group, distinguishing between national grants and self-financed contributions. The average self-financed amount was highest in the non-financial nudge group (1,410,300 JPY), followed by the combination group (936,230 JPY), the control group (757,244 JPY), and the financial incentive group (718,035 JPY). We found statistically significant differences when comparing the non-financial nudge group with each of the combination group ($p < 0.05$), the control group ($p < 0.01$), and the financial incentive group ($p < 0.01$).

## 5.2. Regression Analysis

**[Table 3 is here.]**

**Table 3** presents the results of the regression analysis. In Columns 1 and 2, the dependent variable is the total assessed budget amount, and the treatment variables are binary indicators for the combination group, the financial incentive group, and the non-financial nudge group. In Columns 3 and 4, the dependent variable is the assessed budget amount excluding the national grant. Columns 2 and 4 include additional covariates. We primarily introduce the results with covariates, as the overall patterns remain consistent regardless of covariate inclusion. We used cluster-robust standard errors at the prefecture level to account for the possibility that trends in HPV vaccination awareness campaigns may be correlated within regions.

Columns 1 and 2 show that all three treatments—the combination, the financial incentive, and the non-financial nudge—significantly increased the assessed budget amount compared to the control group. The respective increases were 1,066,810 JPY, 678,350 JPY, and 661,960 JPY, all statistically significant at the 1% level. Moreover, we observed statistically significant differences between the combination group and either the financial incentive group or the non-financial nudge group, while no significant difference emerged between the financial incentive and the non-financial nudge groups. These findings confirm that the assessed budget amounts followed the order: combination group > financial incentive group ≈ non-financial nudge group > control group, consistent with the descriptive statistics. The standardized effect sizes for the combination, financial incentive, and non-financial nudge groups were 0.50, 0.35, and 0.35, respectively. Typically, an effect size of 0.50 is interpreted as moderate, while those around 0.35 are considered small to moderate. The effect size of the non-financial nudge is comparable to that reported by Kuroki and Sasaki (2023a).

In Columns 3 and 4, where the dependent variable is the assessed budget amount excluding the national grant, we found that neither the combination treatment nor the financial incentive significantly increased the self-financed portion of the budget. The differences relative to the control group were 140,470 JPY and –34,350 JPY, respectively—both statistically insignificant and modest in magnitude. These results suggest that while the combination and financial incentive groups did not alter the self-financed amount compared to the control group, the total assessed amount was higher due to the inclusion of the national grant. In contrast, the non-financial nudge led to an increase of 661,960 JPY in the self-financed budget amount. Given that the non-financial nudge alone increased the total assessed amount by a similar magnitude as the financial incentive alone in Columns 1 and 2, this suggests that the increase in the self-financed amount in the non-financial nudge group was equivalent to the subsidy amount that the central government would have provided in the financial incentive group.

Regarding the covariates, we found that years of experience in the respondent's current department and the population size of the municipality were both negatively associated with the assessed budget amount, with coefficients statistically significant at the 10% level or lower. These results imply that longer tenure in the finance department and larger municipal population sizes are associated with more conservative budget assessments.

## 6. Supplementary Analysis
### 6.1. Manipulation Check

We first conducted a manipulation check to verify whether the treatments influenced budget officers' sensitivity to the financial incentive or non-financial nudge. We assessed this by comparing how respondents perceived the funding-related and outcome-related information during the budget assessment process. Sensitivity was measured using a 7-point Likert scale (1 = poor, 7 = excellent).

**[Figure 2 is here.]**

As shown in **Figure 2**, the combination group and the financial incentive group—both of which received information indicating that half of the campaign costs would be covered by national subsidies—exhibited significantly higher sensitivity to the projected funding for FY2024 compared to the control group. The mean sensitivity score was 3.200 in the control group, whereas it was 4.631 in the combination group and 4.833 in the financial incentive group. The non-financial nudge group also showed higher sensitivity (3.610) than the control group, although its mean score was lower than those of the other two treatment groups.

Similarly, the combination group and the non-financial nudge group—both of which received information highlighting the projected outcomes of resuming HPV vaccination recommendations through a loss-framed message—reported significantly higher sensitivity to the performance targets for FY2024. The mean sensitivity score in the control group was 3.959, while it was 4.324 in the combination group and 4.257 in the non-financial nudge group. In contrast, the financial incentive group's sensitivity score was 3.879, which was not statistically different from the control group.

These results suggest that the financial incentive and non-financial nudge interventions increased budget officers' sensitivity to the relevant information, thereby influencing their budget assessments in the expected direction.

**6.2. Individual and Team's Decision-Makings**

In typical budgeting procedures, an individual budget officer conducts the initial assessment, which is subsequently reviewed and adjusted by a manager as part of the secondary assessment. Reflecting this institutional practice, we analyzed how decisions made by individual officers and by managers contributed to the final assessed budget amount. To enable this analysis, our questionnaire included two fields: one for the initial assessment made independently by the officer, and another for the amount determined after consultation with their manager.

**[Figure 3 is here.]**

**Figure 3** presents the results of a mediation analysis on assessed amounts using Structural Equation Modeling (SEM), showing only the statistically significant paths (at the 10% level or lower). The analysis reveals that the initial decisions made by individual budget officers had a substantial influence on the final assessed amounts. Specifically, for every 1 JPY increase in the initial assessment, the amount after managerial review increased by 0.867 JPY, indicating that most of the final assessed amounts could be explained by the officers' original entries. We also confirmed that the effects of each treatment on individual assessments were generally consistent with the results reported in Section 5.2.

Among all treatment groups, only the financial incentive group exhibited a statistically significant increase in the assessed amount after consultation with managers, with an average increase of 183,390 JPY relative to the initial individual assessment. This result suggests that managers may have deemed it appropriate to raise the assessed amount slightly, considering that half of the project cost would be covered by national subsidies. In contrast, the combination group—which also received the financial incentive—did not show any additional increase after managerial review, possibly because the initial assessments were already regarded as appropriate.

We also conducted a mediation analysis for the probability of assigning a non-zero assessed amount (see **Appendix 2**). This analysis showed that in both the financial incentive group and the non-financial nudge group, the likelihood of reporting a non-zero assessment increased after the managerial review. These results suggest that in the single-intervention groups, initial assessments by individual officers may have been conservative, and that managers subsequently revised the amounts upward.

Taken together, these findings indicate that in real-world budget assessment settings, the initial judgments of individual budget officers could substantially shape the final outcomes, with managerial reviews potentially making further adjustments depending on the intervention.

**6.3. Appropriateness of Treatment-Induced Budget Increases**

We found that all three treatments—the financial incentive, the non-financial nudge, and their

combination—increased the assessed budget amounts provided by respondents. A natural follow-up question is whether these treatments brought the assessed amounts closer to a more appropriate level. To examine this, we divided the analytical sample into two subgroups: respondents with a myopic tendency to discount future outcomes and those with a far-sighted tendency to evaluate future outcomes without discounting.

Our questionnaire included a behavioral economics question designed to measure time discounting. Respondents were asked whether they would prefer (1) to receive 100,000 JPY today or (2) to receive 110,000 JPY one year later, under the assumption that the payment was guaranteed. Among the 490 respondents, 306 chose (1) or leaned toward it, indicating that they discounted future outcomes and could be classified as the myopic group. In contrast, 184 chose (2) or leaned toward it, suggesting that they evaluated future outcomes at face value and could be classified as the far-sighted group.

If the upward effects of the treatments are observed primarily in the myopic group, we may interpret this as an adjustment that corrects downward assessments resulting from time discounting. Conversely, if the treatment effects appear predominantly in the far-sighted group, it could suggest that the treatments push assessed amounts beyond necessary levels.

**[Table 4 is here.]**

The estimation results in **Table 4** show that the increases in assessed amounts from the combination treatment, the financial incentive alone, and the non-financial nudge alone were primarily observed within the myopic group. The estimated effects were 1,369,192 JPY ($p < .01$), 787,776 JPY ($p < .01$), and 878,923 JPY ($p < .01$), respectively. In contrast, none of the treatments produced statistically significant effects within the far-sighted group.

The constant terms indicate that, in the control group, the average assessed amount was 497,367 JPY for the myopic group and 1,265,674 JPY for the far-sighted group. If we consider the value for the far-sighted group to represent a more appropriate assessment level unaffected by time discounting, the increase observed in the myopic group as a result of the treatments can be interpreted as bringing their assessments closer to that level. In particular, in the financial incentive and non-financial nudge groups, the sums of the constant term and the estimated treatment effects were 1,285,143 JPY and 1,376,290 JPY, respectively—nearly identical to the far-sighted group's constant term.

Because the effect of the combination treatment was even larger, the assessed amount in this group reached 1,866,559 JPY, exceeding the benchmark level represented by the far-sighted group. This could suggest that the combination treatment may have led to overestimation. However, given that the original requested amount was 5,000,000 JPY, it is also possible that the appropriate assessment level exceeds the far-sighted group's constant term. Moreover, our time discounting measure was based on a one-year comparison; respondents who are willing to wait one year may also be willing to wait two or three years, and such individuals might report even higher assessed budget amounts. Under this interpretation, the combination treatment may also have contributed to moving the assessed amounts closer to a more appropriate level.

**7. Discussion and Conclusions**

This study yielded two primary findings. First, among all treatments, the combination treatment produced the largest increase in the total assessed budget amount, followed by the financial incentive and non-financial nudge treatments, both of which had comparable effects. Second, only the non-financial nudge treatment significantly increased the self-financed portion of the budget, whereas the combination and financial incentive treatments did not differ from the control group in this regard. In determining the volume of public services to be delivered to citizens, the total budget size is critical. At the same time, from the perspective of revenue allocation and fiscal autonomy, the size of the self-financed budget is also important (Anessi-Pessina et al., 2016).

Our findings suggest that when the objective is to increase the total assessed budget amount—regardless of the funding source—it is effective to provide both incentive-based and nudge-based information during the assessment request process. The fact that the combination treatment led to a greater increase in budget assessments than either treatment alone indicates that budget officers are heterogeneous: some respond more strongly to financial information such as incentives, while others are more influenced by non-financial cues such as nudges. By presenting both types of information simultaneously, the combination treatment likely affects both types of decision-makers.

Furthermore, our results indicate that when the goal is to increase the assessed budget amount under the constraint that it must be funded by local self-financed revenue, providing only nudge-based information is more effective. Although the combination treatment included the same nudge content, it did not lead to an increase in the self-financed amount. Drawing on insights from behavioral science (Frey & Jegen, 2021), it is possible that the nudge worked to strengthen respondents' intrinsic motivation to allocate local resources, whereas the presence of financial incentives may have crowded out that intrinsic motivation.

Of course, this study has limitations. The most salient limitation is the uncertain external validity of our findings, given that the experiment relied on hypothetical scenarios. As discussed in **Section 4.2**, we made several efforts to enhance the realism of the scenarios and verified their plausibility through interviews with practitioners. Moreover, we confirmed that the treatment effects reported above were also observed among respondents who worked in municipalities with population sizes similar to that described in the hypothetical scenario. The finding that information about other municipalities increased assessed budget amounts is consistent with previous research demonstrating interdependence among local governments (Case et al., 1993; Walker, 1969). Nevertheless, the generalizability of our findings should be tested in future studies.

Budget assessment decision-making has long been considered a black box. Most previous studies have relied on descriptive surveys asking what kinds of information are used in assessments or on correlation analyses examining relationships between assessed amounts and the existence of processes or output indicators (Liguori et al., 2012; Marti, 2013; Melkers and Willoughby, 2001, 2005; Zaltsman, 2009). Despite the limitations in generalizability, our study makes a novel contribution by identifying the causal effects of treatments—such as financial incentives and nudges—on previously unexplored decision-making processes.

**References**

Anessi-Pessina, E., C. Barbera, M. Sicilia, and I. Steccolini. 2016. Public Sector Budgeting: A European Review of Accounting and Public Management Journals. *Accounting, Auditing & Accountability Journal.* **29**(3): 491-519.

Baekgaard, M. 2015. Performance Information and Citizen Service Attitudes: Do Cost Information and Service Use Affect the Relationship?, *International Public Management Journal.* **18**(2), 228-245. https://doi.org/10.1080/10967494.2015.1022676

Case, Anne C., Harvey S. Rosen, and James R. Hines. 1993. Budget Spillovers and Fiscal Policy Interdependence. *Journal of Public Economics* **52**(3): 285–307. https://doi.org/10.1016/0047-2727(93)90036-S

DellaVigna, S., & Linos, E. (2022). RCTs to scale: Comprehensive evidence from two nudge units. *Econometrica,* **90**(1), 81-116.

Demaj, L. 2017. What Can Performance Information Do to Legislators? A Budget-Decision Experiment with Legislators, *Public Administration Review*. **77**(3): 366-379.

Druckman, J. N., D. P. Green, J. H. Kuklinski, and A. Lupia. 2014. "Experimentation in Political Science." In Cambridge Handbook of Experimental Political Science, edited by J. N. Druckman, D. P. Green, J. H. Kuklinski, and A. Lupia, 3-11. Cambridge University Press.

Epperson, R., & Reif, C. (2019). Matching subsidies and voluntary contributions: A review. *Journal of Economic Surveys,* **33**(5), 1578-1601.

Frey, B. S., & Jegen, R. (2001). Motivation crowding theory. *Journal of Economic Surveys*, **15**(5), 589-611.

Gilmour, J.B. and Lewis, D.E. 2006. Does Performance Budgeting Work? An Examination of the Office of Management and Budget's PART Scores. *Public Administration Review.* **66**: 742-752. https://doi.org/10.1111/j.1540-6210.2006.00639.x

Heinrich, C. J. 2012. How Credible Is the Evidence, and Does It Matter? An Analysis of the Program Assessment Rating Tool. *Public Administration Review.* **72**(1): 123-34. https://doi.org/10.1111/j.1540-6210.2011.02490.x

Ikeda, S., Y. Ueda, A. Yagi, S. Matsuzaki, and E. Kobayashi, 2019. HPV vaccination in Japan: what is

happening in Japan?. *Expert Review of Vaccines,* **18**(4): 323-325, DOI: 10.1080/14760584.2019.1584040

Jacobs A. M. 2011. *Governing for the Long Term: Democracy and the Politics of Investment.* New York: Cambridge Univ. Press.

Jacobs, A. M. 2016. Policy Making for the Long Term in Advanced Democracies. *Annual Review of Political Science.* **19**:1, 433-454.

Karlan, D., & List, J. A. (2007). Does price matter in charitable giving? Evidence from a large-scale natural field experiment. *American Economic Review,* **97**(5), 1774-1793.

Kuroki, M. (2022). Impact of Depreciation Information on Capital Budgeting among Local Governments: A Survey Experiment. *Australian Accounting Review.* https://doi.org/10.1111/auar.12355

Kuroki, M. and K. Motokawa. (2022). Do Non-financial Performance and Accrual-Based Cost Information Affect Public Sector Budgeting?, *Journal of Public Budgeting, Accounting & Financial Management.* **34**(6): 95-116. https://doi.org/10.1108/JPBAFM-03-2021-0056

Kuroki, M. and S. Sasaki. (2023a). Nudging Public Budget Officers: A Field-Based Survey Experiment. *Public Budgeting & Finance* **43**: 3–20. https://doi.org/10.1111/pbaf.12345

Kuroki, M. and S. Sasaki. (2023b). Applying Behavioral Science to Budget Requests and Budget Assessments in Japanese Local Governments. AEA RCT Registry. https://doi.org/10.1257/rct.12438-1.0

Liguori, M., Sicilia, M. and Steccolini, I. (2012), Some like it non-financial politicians' and managers' views on the importance of performance information, *Public Management Review,* Vol. 14 No. 7, pp. 903-922.

Loewenstein, G., D. Read, and R. Baumeister. (2003). *Time and Decision: Economic and Psychological Perspectives on Intertemporal Choice*. Russell Sage Foundation.

Martí, C. (2013), Performance budgeting and accrual budgeting, *Public Performance and Management Review,* Vol. 37 No. 1, pp. 33-58.

Mauro, S. G. (2021). *Budgeting and performance management in the public sector*. Routledge.

McGowan, M. J., J. E. V. Pope, M. E. Kropf, and Z. Mohr. (2021). Guns or Butter… or Elections? Understanding Intertemporal and Distributive Dimensions of Policy Choice Through the Examination of Budgetary Tradeoffs at the Local Level. *Public Budgeting & Finance* **41**: 3– 19. https://doi.org/10.1111/pbaf.12289

Melkers, J. and Willoughby, K. (2001), Budgeters' views of state performance-budgeting systems: distinctions across branches, *Public Administration Review,* Vol. 61 No. 1, pp. 54-64.

Melkers, J. and Willoughby, K. (2005), Models of performance-measurement use in local governments: understanding budgeting, communication, and lasting effects, *Public Administration Review,* Vol. 65, pp. 180-190.

Ministry of Health, Labour and Welfare HPV Vaccine Website. https://www.mhlw.go.jp/bunya/kenkou/kekkaku-kansenshou28/index.html (last accessed July 20, 2024)

Mohr, Z., and L. Kearney. (2021). Behavioral-Experimental Public Budgeting and Financial Management: A Review of Experimental Studies in the Field. *Public Finance and Management.* **20**(1).

Namba, M., Y. Kaneda, C. Kawasaki, R. Shrestha, and T. Tanimoto. (2023). Underlying background of the current trend of increasing HPV vaccination coverage in Japan. *Global Health & Medicine.* 2023; 5(4):255-256. DOI: 10.35772/ghm.2023.01010

Nielsen, P. A. and M. Baekgaard. (2015). Performance Information, Blame Avoidance, and Politicians' Attitudes to Spending and Reform: Evidence from an Experiment. *Journal of Public Administration Research and Theory.* **25**(2): 545–569.

Reck, J.L. (2001). The usefulness of financial and nonfinancial performance information in resource allocation decisions. *Journal of Accounting and Public Policy.* **20**(1): 45-71.

Rubin, Irene. (2008). *Public Budgeting: Policy, Process, and Politics*. New York: ME Sharpe Inc.

Simon, Herbert A. (1997). *Administrative Behavior: A Study of Decision-Making Processes in Administrative Organizations* (4th ed.). Free Press.

Sunstein, C. R. (2014). *Why nudge?: The politics of libertarian paternalism*. Yale University Press.

Thaler, R. H. and C.R. Sunstein. (2009). *Nudge: Improving Decisions about Health, Wealth, and Happiness.* New York: Penguin Books.

Thurmaier, K. (1992). Budgetary Decision Making in Central Budget Bureaus: An Experiment. *Journal of Public Administration Research and Theory.* 2: 463-487. https://www.jstor.org/stable/1181702

Thurmaier, K. (1995). Decisive Decision Making in the Executive Budget Process: Analyzing the Political and Economic Propensities of Central Budget Bureau Analysts. *Public Administration Review.* **55**(5): 448-460. https://doi.org/10.2307/976769

Tversky, A. and D. Kahneman. (1981). The Framing of Decisions and the Psychology of Choice. *Science.* **211**(4481): 453-458.

Tversky, A. and D. Kahneman. (1991). Loss Aversion in Riskless Choice: A Reference-Dependent Model. *The Quarterly Journal of Economics.* **106**(4): 1039-1061. https://doi.org/10.2307/2937956

Tversky, A. and D. Kahneman. (1992). Advances in Prospect Theory: Cumulative Representation of Uncertainty. *Journal of Risk and Uncertainty.* **5**: 297-323. https://doi.org/10.1007/BF00122574

Walker, Jack L. (1969). The diffusion of innovations among the American states. *American Political Science Review* **63**(3): 880–899. https://doi.org/10.2307/1954434

Weingast, B. R. (2009). Second generation fiscal federalism: The implications of fiscal incentives. *Journal of Urban Economics.* **65**(3): 279-293.

Zaltsman, A. (2009). The Effects of Performance Information on Public Resource Allocations: A Study of Chile's Performance-Based Budgeting System. *International Public Management Journal.* **12**(4): 450-483. https://doi.org/10.1080/10967490903328931

**Table 1. Response Distributions**

Panel A: Survey Targets

|  | Incentive × Nudge | Incentive | Nudge | Control | Total |
|---|---|---|---|---|---|
| Tokyo Metropolitan | 6 | 6 | 6 | 5 | 23 |
| Designated City | 5 | 5 | 5 | 5 | 20 |
| Ordinary City | 193 | 193 | 193 | 193 | 772 |
| Towns and Villages | 232 | 231 | 231 | 232 | 926 |
| Total | 436 | 435 | 435 | 435 | 1741 |

Panel B: Survey Responses

|  | Incentive × Nudge | Incentive | Nudge | Control | Total |
|---|---|---|---|---|---|
| Tokyo Metropolitan | 1 | 0 | 1 | 1 | 3 |
| Designated City | 1 | 3 | 2 | 1 | 7 |
| Ordinary City | 47 | 45 | 57 | 63 | 212 |
| Towns and Villages | 69 | 66 | 62 | 71 | 268 |
| Total | 118 | 114 | 122 | 136 | 490 |

$$\chi^2 = 5.195$$

**Table 2. Descriptive Statistics**

**Panel A: Descriptive Statistics of Total Sample**

| Variables | mean | s.d. | min | 25% | median | 75% | max |
|---|---|---|---|---|---|---|---|
| *Total budget amount* | 134.633 | 193.489 | 0.000 | 0.000 | 20.000 | 200.000 | 1000.000 |
| *Incentive × Nudge* | 0.241 | 0.428 | 0.000 | 0.000 | 0.000 | 0.000 | 1.000 |
| *Incentive* | 0.233 | 0.423 | 0.000 | 0.000 | 0.000 | 0.000 | 1.000 |
| *Nudge* | 0.249 | 0.433 | 0.000 | 0.000 | 0.000 | 0.000 | 1.000 |
| *Gender* | 0.869 | 0.337 | 0.000 | 1.000 | 1.000 | 1.000 | 1.000 |
| *Experience* | 3.394 | 2.589 | 1.000 | 2.000 | 3.000 | 4.000 | 30.000 |
| *Age* | 2.349 | 0.788 | 1.000 | 2.000 | 2.000 | 3.000 | 4.000 |
| *Population* | 9.947 | 1.496 | 5.790 | 8.881 | 9.912 | 10.889 | 15.145 |
| *12 to 16 Ratio (%)* | 4.017 | 0.751 | 0.610 | 3.624 | 4.016 | 4.453 | 7.039 |
| *District* | 6.193 | 1.296 | 2.342 | 5.385 | 6.101 | 6.916 | 9.806 |
| *Immigrants Ratio (%)* | -0.464 | 0.999 | -12.520 | -0.800 | -0.450 | 0.010 | 2.290 |
| *Income* | 2870.761 | 428.057 | 2131.400 | 2576.300 | 2785.900 | 3095.900 | 4706.400 |

**Panel B: Means of Each Group**

| Variables | *Incentive × Nudge* | *Incentive* | *Nudge* | *Control* | Total |
|---|---|---|---|---|---|
| *Total budget amount* | 187.246 | 143.607 | 141.030 | 75.724 | 134.633 |
| *Budget_amount (dummy)* | 80.508% | 64.912% | 62.295% | 48.529% | 63.469% |
| *Gender* | 0.898 | 0.877 | 0.861 | 0.846 | 0.869 |
| *Experience* | 3.475 | 3.500 | 3.270 | 3.346 | 3.394 |
| *Age* | 2.466 | 2.228 | 2.393 | 2.309 | 2.349 |
| *Population* | 9.962 | 9.828 | 9.964 | 10.019 | 9.947 |
| *12 to 16 Ratio (%)* | 3.990 | 3.994 | 4.016 | 4.061 | 4.017 |
| *District* | 6.217 | 6.118 | 6.161 | 6.262 | 6.193 |
| *Immigrants Ratio (%)* | -0.545 | -0.544 | -0.435 | -0.351 | -0.464 |
| *Income* | 2,895.736 | 2,857.687 | 2,821.664 | 2,904.094 | 2,870.761 |

*Note*: N=490.

**Table 3. Regression Results**

| Variables | Total budget amount | | Total budget amount – national grant | |
|---|---|---|---|---|
| | (1) | (2) | (3) | (4) |
| Constant | 75.724 *** | 141.128 | 75.724 *** | 99.394 |
| | (14.259) | (86.916) | (14.259) | (62.781) |
| Incentive × Nudge group | 111.522 *** | 106.681 *** | 17.899 | 14.047 |
| | (29.728) | (29.114) | (19.367) | (19.700) |
| Incentive group | 67.883 *** | 67.835 *** | -3.920 | -3.435 |
| | (22.562) | (22.292) | (16.629) | (15.892) |
| Nudge group | 65.306 *** | 66.196 *** | 65.306 *** | 66.538 *** |
| | (22.276) | (22.542) | (22.276) | (22.440) |
| Gender | | 31.601 | | 25.702 |
| | | (22.319) | | (17.767) |
| Experience | | -4.810 * | | -3.761 * |
| | | (2.585) | | (2.001) |
| Age | | 17.909 | | 16.470 |
| | | (12.470) | | (10.500) |
| Population | | -23.027 *** | | -20.384 ** |
| | | (8.146) | | (6.431) |
| 12 to 16 ratio | | -20.640 | | -10.941 |
| | | (14.626) | | (11.109) |
| District | | 8.308 | | 7.024 |
| | | (12.448) | | (8.732) |
| ImigrantsRatio | | 5.165 | | 4.325 |
| | | (6.500) | | (5.179) |
| Income | | 0.051 | | 0.047* |
| | | (0.035) | | (0.025) |
| N | 490 | 490 | 490 | 490 |
| adj. $R^2$ | 0.038 | 0.055 | 0.028 | 0.050 |

*Notes:* *, **, and *** indicate statistical significance at the 10%, 5%, and 1% levels, respectively. We used cluster-robust standard errors at the prefecture level.

**Table 4. Treatment Effects by Time Discounting Tendency**

| Dependent variable: | Myopic group | | Far-sighted group | |
|---|---|---|---|---|
| *Total budget amount* | (1) | | (2) | |
| Constant | 49.737 | *** | 126.567 | *** |
|  | (13.313) | | (31.750) | |
| *Incentive × Nudge group* | 136.919 | *** | 61.786 | |
|  | (35.491) | | (41.246) | |
| *Incentive group* | 78.778 | *** | 41.051 | |
|  | (26.896) | | (45.448) | |
| *Nudge group* | 87.892 | *** | 18.889 | |
|  | (24.630) | | (32.105) | |
| N | 306 | | 184 | |
| adj. R-sq | 0.075 | | 0.013 | |

*Notes:* *, **, and *** indicate statistical significance at the 10%, 5%, and 1% levels, respectively. We used cluster-robust standard errors at the prefecture level.

**Figure 1. Assessed Budget Amounts by Group: National Grant and Self-financed Contributions**

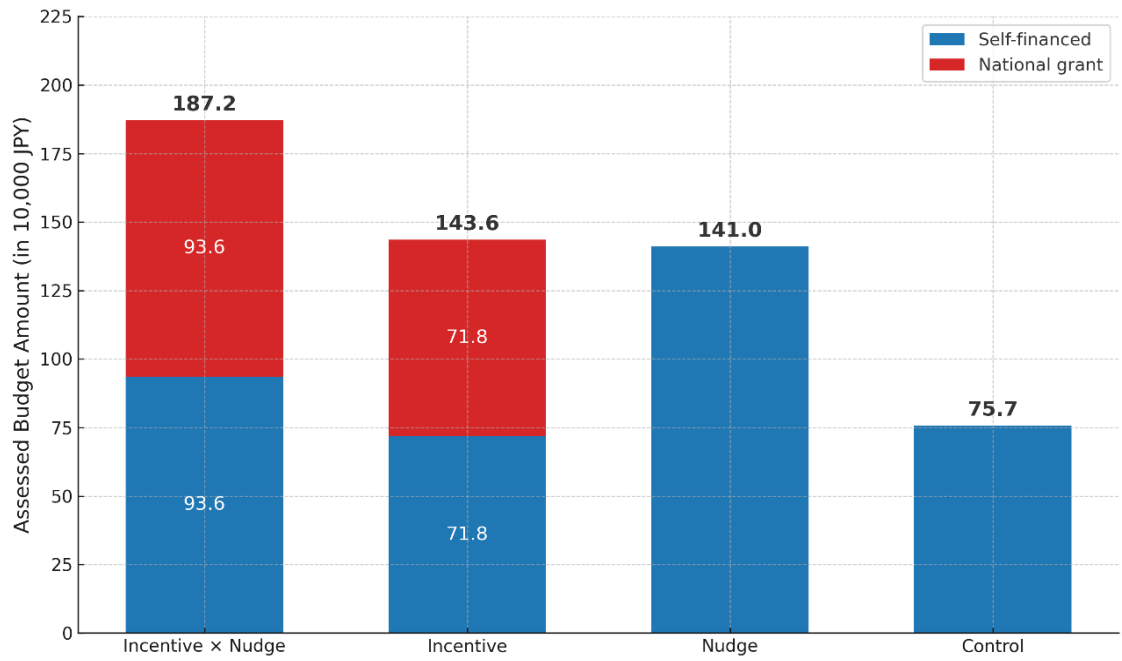

*Notes*: N=490.

**Figure 2. Supplementary Analysis: Manipulation Check**

**Panel A: Sensitivity to Financial Resources**

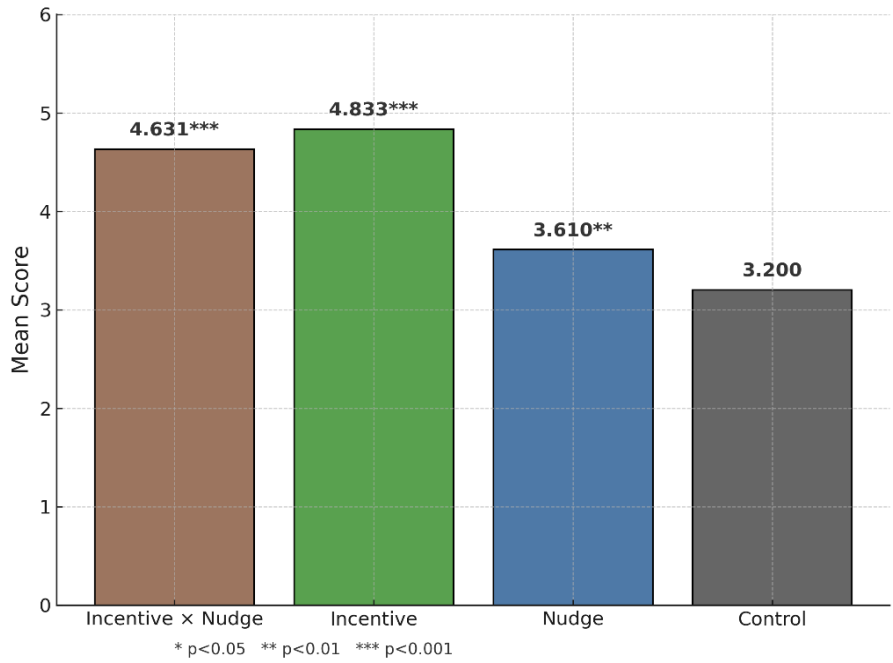

**Panel B: Sensitivity to Performance Targets**

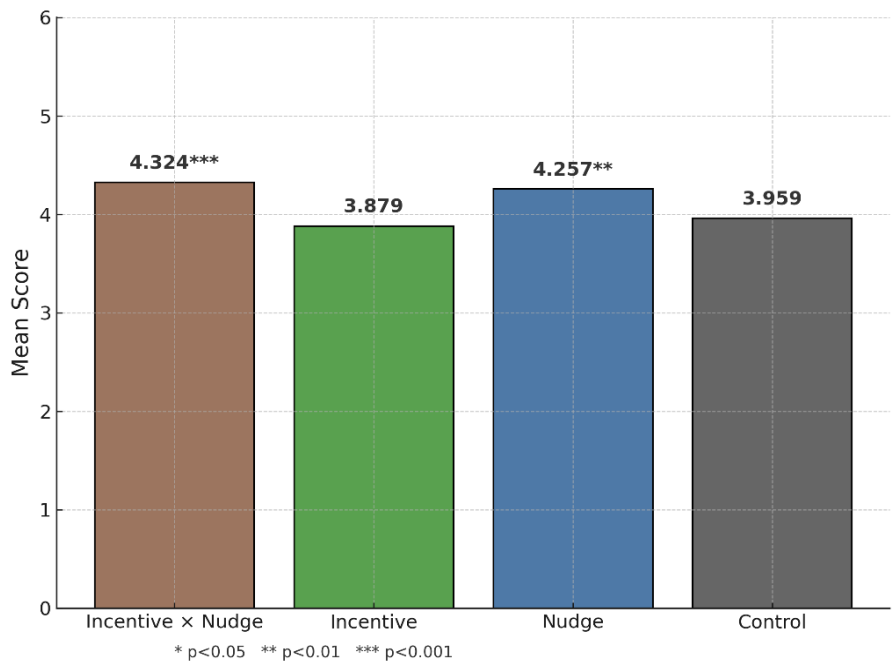

*Notes*: N=490. *, **, and *** indicate statistical significance at the 10%, 5%, and 1% levels, respectively, with the comparison to the control group. We used cluster-robust standard errors at the prefecture level

**Figure 3. Individual and Team's Decision-Makings (Total Budget Amount)**

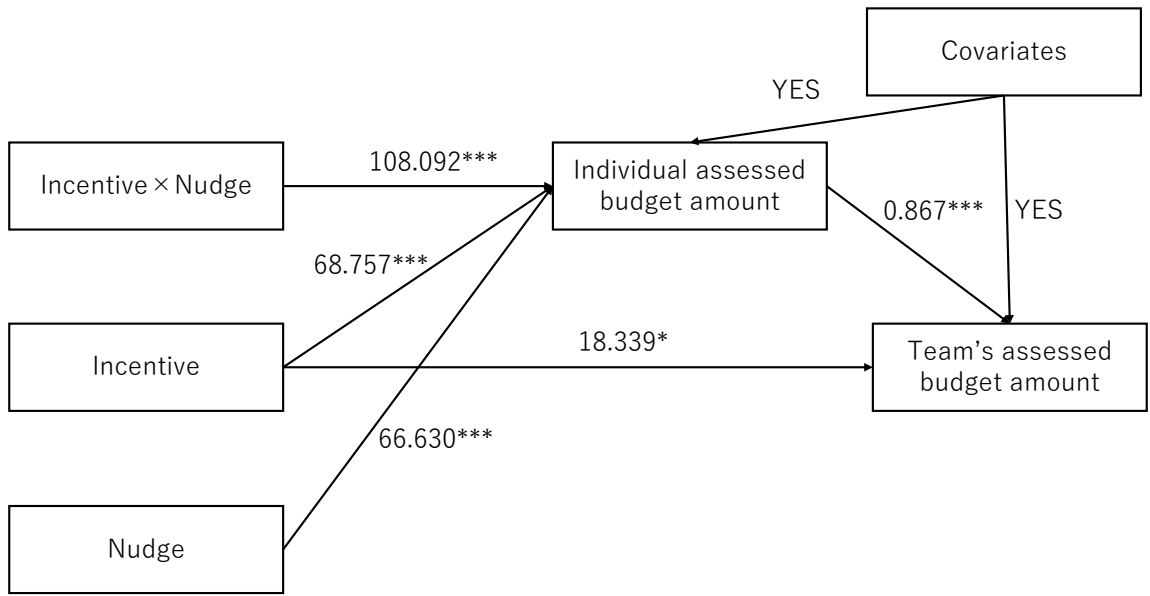

*Notes*: *, **, and *** indicate statistical significance at the 10%, 5%, and 1% levels, respectively. We used cluster-robust standard errors at the prefecture level.

**APPENDIX 1: Content of Survey (in Japanese)**

以下は、<u>あなたが 5 万人程度の自治体の予算編成担当部局に勤めていることを前提</u>にお答えください。自治体における予算機能を明らかにする重要な設問です。正解はなく、個別の回答情報は公開いたしません。<u>設問をよく読んでいただき、次のページのご回答にご協力ください。</u>

※HPV ワクチン普及啓発事業をテーマとしていますが、周知啓発段階の事業をいくつか議論の上で選択しました。HPV ワクチンそのものの是非について問う設問ではございません。

① HPV ワクチンの受診勧奨に関する啓発事業の予算要求書

本自治体では人口減少によって財源が厳しい状況で<u>前年度を上回らない金額での予算要求を推奨</u>しています。健康関連部局の担当者から下記のとおり、2023 年度は本事業についてゼロ査定であった 2024 年度予算要求が上がりました。

金額欄に各事業仮の査定金額について、<u>次のページに回答者個人及び課長級等の責任者と相談した上での 2 つの査定額でお答えください</u>。

ベースライン情報

<参考資料>

| 予算額 | 2024 年度要求額 | 2023 年度 | 2022 年度 |
|---|---|---|---|
| 啓発事業 | 500 万円 | 0 万円 | ― |

<事業の概要>

> HPV ワクチン（ヒトパピローマウイルス様粒子ワクチン）は、2022 年度の厚生労働省の安全対策調査会で、安全性について特段の懸念はなく、接種の有効性が副反応のリスクを明らかに上回ることが認められました。
>
> 日本では、25～40 歳の女性のがんによる死亡の第 2 位が子宮頸がんによるものですが、HPV ワクチンの接種で 10 万人当り 209～144 人が子宮頸がんによる死亡を回避できると報告されています。
>
> （参考資料 https://www.mhlw.go.jp/stf/seisakunitsuite/bunya/kenkou/hpv_9-valentHPVvaccine.html）
>
> 本市の定期接種の対象年代である 14～16 歳の接種率は、「5％」と著しく低い状況です（2023 年度見込み）。そこで 2024 年度には、予防接種法第 8 条の規程による積極勧奨として、啓発効果の高い独自のチラシを製作して、対象世代の約 500 名に配布する事業を行う予定です。
>
> *XXX・・・*

<成果情報>

| 成果 | 2024 年度成果目標 | 2023 年度接種率見込み | 2022 年度接種率 | 2021 年度接種率 |
|---|---|---|---|---|
| 接種促進事業全体 | 20%<br>100 人 | 5%<br>25 人 | 2%<br>10 人 | 1%<br>5 人 |

E1-1 査定金額（回答欄）

| 事業名 | 個人の査定額 | 課長等と相談後の査定額 |
|---|---|---|
| HPV ワクチン啓発事業 | 万円 | 万円 |

※国からの補助額を含めた総額を記入してください。*該当する介入群にのみ記載*

E1-2 なぜ上記の金額に査定したのかについて自由に記載をお願いします。

E1-3 事業の成果情報についてどのように評価しますか？（○を付けてください）。また、それぞれの情報について参考にした割合を記載してください。

| 項目 | 悪い | | どちらでもない | | | 優れている | |
|---|---|---|---|---|---|---|---|
| 2022 年度以前の接種率 | 1 | 2 | 3 | 4 | 5 | 6 | 7 |
| 2023 年度の接種率 | 1 | 2 | 3 | 4 | 5 | 6 | 7 |
| 2024 年度の成果目標 | 1 | 2 | 3 | 4 | 5 | 6 | 7 |
| 2023 年度の査定結果 | 1 | 2 | 3 | 4 | 5 | 6 | 7 |
| 2024 年度の予定財源 | 1 | 2 | 3 | 4 | 5 | 6 | 7 |

E1-4 主な回答者についてお答えください。

| 部署名 | | 役職名 | 職員 ・ 係長級 ・ 課長級以上 | 性別 | |
|---|---|---|---|---|---|
| 年齢（○をつけてください） | | 20 代 ・ 30 代 ・ 40 代 ・ 50 代 ・ 60 代 | | | |

*Notes*:

**統制群**

　*XXX・・・* 無し

**財務的インセンティブ群**

　*XXX・・・* 今年度、本啓発の財源については、国から「半額の補助」が提供されます。

**非財務的ナッジ群**

　*XXX・・・* 周辺自治体はすでに同様の啓発事業を展開しており、2023 年度の接種率は「20%」に到達する見込みで、社会から注目されています。啓発事業を行わない場合、本市の接種率は低いまま推移すると予想されます。

**組み合わせ群**

　*XXX・・・* 今年度、本啓発の財源については、国から「半額の補助」が提供されます。周辺自治体はすでに同様の啓発事業を展開しており、2023 年度の接種率は「20%」に到達する見込みで、社会から注目されています。啓発事業を行わない場合、本市の接種率は低いまま推移すると予想されます。

**APPENDIX 2: Individual and Team's Decision-Makings (Non-zero Budget Assessment)**

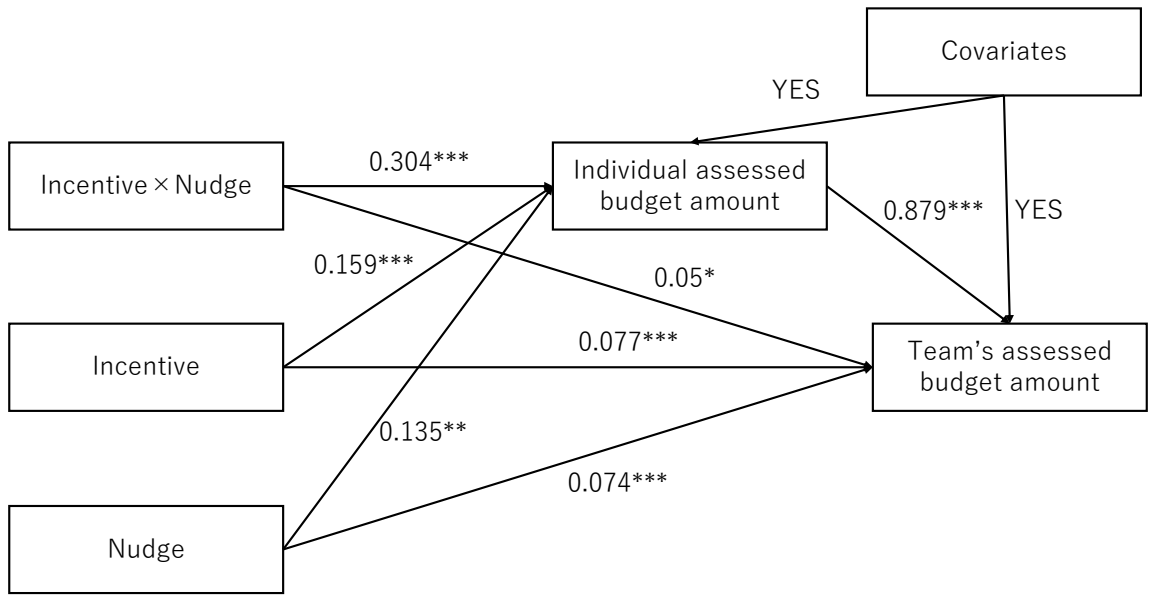

*Notes*: *, **, and *** indicate statistical significance at the 10%, 5%, and 1% levels, respectively. We used a linear probability model and cluster-robust standard errors at the prefecture level.